\documentclass[12pt,a4paper,final]{iopart}

\usepackage{iopams}  
\usepackage{graphicx}
\usepackage{epstopdf}
\usepackage{textcomp}
\usepackage[breaklinks=true,colorlinks=true,linkcolor=blue,urlcolor=blue,citecolor=blue]{hyperref}
\begin{document}

\title[]{Spectroscopic Ellipsometry investigation on anisotropic optical properties of sputtered AlN films}

\author{Padmalochan Panda$^a$, R. Ramaseshan$^{a*}$, S. Tripura Sundari$^{a*}$ and H. Suematsu$^b$}
\address{$^a$Surface and Nanoscience Division, Materials Science Group, Indira Gandhi Centre for Atomic Research, HBNI, Kalpakkam - 603102, India.}
\address{$^b$Extreme Energy Density Research Institute, Nagaoka University of Technology, Nagaoka, Japan.}

\ead{seshan@igcar.gov.in}
\ead{sundari@igcar.gov.in}

\begin{abstract}
We report the uniaxial anisotropic optical constants of Wurtzite type AlN films deposited on Si (100) substrate using DC reactive magnetron sputtering as a function of growth temperature ($T_s$, 35 to 600 $^\circ$C). Evolution of optical properties with $T_s$ is investigated by Spectroscopic Ellipsometry (SE) technique. Thickness and roughness of these films are determined from the regression analysis of SE data, which are corroborated using TEM and AFM technique. Highly a-axis oriented AlN film grown at 400 $^\circ$C, exhibits high \emph{n} and low \emph{k} at 210 nm (deep-UV region) with a small birefringence and dichroism neat to band edge, which can be used in isotropic deep UV opto-electronic device applications. All these AlN films exhibit transparent nature from near infrared (NIR) to 354 nm, where optical band gap energies vary between 5.7 to 6.1 eV.
\end{abstract}



In the last two decades, aluminum nitride (AlN) films have been attracted extensively in the semiconductor industry due to their unique outstanding physical and optical properties with great technological advantages. AlN is a wide band gap ($\thicksim$ 6.2 eV) semiconductor in addition to its high thermal conductivity and low thermal co-efficient of expansion with high breakdown dielectric strength \cite{Hadis, Pan}. Therefore, It is used in ultraviolet polarizer, optoelectronic displays, high temperature devices and short wavelength light source/detector applications \cite{Hadis, Mustafa, Taniyasu}. AlN and Al-rich AlGaN alloy films find a crucial role in UV-LED (covering wavelengths from 210 to 375 nm), where the insulating properties can be used in the fabrication of GaN, GaAs and InP based electronic, radio-frequency UV sensor and optical devices \cite{Chivukula, Aloki}. Since they have high figure of merit for piezoelectric response, they are suitable for ultra-small mechanical pressures. \cite{Bishara}. They are potential candidates to replace Lead zirconate titanates, since they have high dielectric strength, ease of deposition (involving low temperatures and nontoxic precursors). 

Generally, properties of these films depend on the crystal structure, crystal orientation and micro-structure that in turn depend on the deposition conditions. Due to simplicity and reproducibility, reactive magnetron sputtering technique is one of the common methods for growing AlN films under low temperature with desired crystal structure and orientation \cite{Bishara, Feby, Galca, Medjani, Kuan}. For wurzite AlN structure, the top of valence band creates excitonic states due to the hexagonal crystal-field and spin–orbital splittings \cite{Chen, Silveira}. So, there are two types configuration of excitonic transitions \emph{i.e.} $\sigma$ ($E \bot c, c $=axis of wurzite structure) and $\pi$ ($E \| c$) configuration. For, C-plane AlN based deep-ultraviolet light-emitting diodes (deep UV-LED), the near band-edge emission is intrinsically weak along the normal of this plane due to the strong polarization effect (E$\parallel$c). Whereas, A or M-plane deep UV-LED shows 25 times higher isotropic emission intensity along the surface normal compared with the conventional C-plane LED structure \cite{Taniyasu, Kasu}. So, plane orientation heavily affects the radiation properties of UV-LED, polarizer, electro-luminescent diodes like optoelectronic devices. This suggests that it is salient to study and understand both the structural and optical properties of AlN films over a broad range of wavelengths starting from deep UV to NIR to extend and improve the device performance. Growth temperature ($T_s$) of thin films has a crucial role on the micro-structure and orientation, that affects the physical properties including optical and mechanical. Very few reports are available on the optical properties of crystaline AlN thin films with $T_s$ by considering as isotropic optical response \cite{Mustafa, Mahmood}. Spectroscopic Ellipsometry (SE) measurement is a powerful unique optical characterization technique to determine the optical properties of materials in the broad range of energy, which is a non invasive, non-contact and sensitive technique with a high degree of accuracy \cite{Sundari}. So, in this article, we are primarily concerned with the anisotropic optical properties of AlN films deposited on Si (100) substrate by reactive sputtering technique at different $T_s$ and the anisotropic behavior of a-axis oriented AlN film in the photon energy range of 0.6 to 6.5 eV using SE technique.

\begin{figure}[h]
\centering
\includegraphics[width=0.3\textwidth]{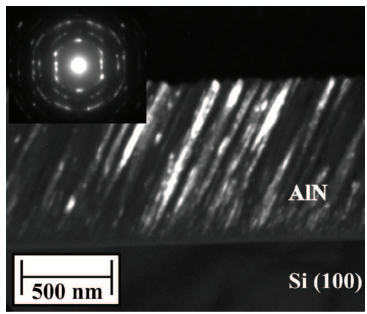}
\caption{Dark field x-TEM and SAED (inset) of AlN films grown at 400 $^\circ$C.}
\label{fig:1}
\end{figure}

AlN thin films were synthesized by DC reactive magnetron sputtering technique (M/s. MECA 2000, France), using a 4N pure aluminum (Al) target, in a mixture of high pure argon (5N) and nitrogen (5N) gases on Si(100) substrate. The flow ratio of sputtering gas (Ar) to reactive gas ($N_2$) was kept constant at 4:1 SCCM with deposition pressure of $5 \times 10 ^{-3}$ mbar for all these films. The target to substrate distance during deposition was maintained at 14 cm. A thin layer of Al was deposited for few seconds to increase the adhesive strength between substrate and deposited AlN films \cite{Panda}. These films were grown at different $T_s$ such as 35, 200, 400 and 600 $^\circ$C. The crystal structure and microstructure of these films were studied by GIXRD and TEM, respectively. All these films are found to be poly-crystalline in nature with hexagonal wurtzite structure except 400 $^\circ$C (highly a-axis oriented) \cite{Panda}. Fig.\ref{fig:1} shows a dark field cross-sectional TEM (x-TEM) image of the film grown at 400 $^\circ$C along with the selected area diffraction (SAED) image in the inset. This shows that the film is highly oriented along a-axis with a slanted columnar structure by making few degrees to the substrate normal as described elsewhere \cite{Panda}.
 
The SE parameters are measured in ambient conditions by a phase modulated spectroscopic ellipsometer (M/s. Horiba Jobin-Yvon, UVISEL2, France) at an incident angle of 70$^\circ$ in the photon energy range of 0.6 to 6.5 eV with 0.01 eV increment. SE measures change in polarization state of light as it reflects from the sample of interest. The polarization change between the parallel (\textit{p}) and perpendicular (\textit{s}) components of the reflected light with respect to the plane of incidence is represented as the change in amplitude ($\Psi$) and the phase difference ($\Delta$), which are considered as ellipsometric parameters. The ellipsometric parameters $\Psi$ and $\Delta$ are defined by the Eq. (1) below. 
 
\begin{equation}
\rho = \frac{r_p}{r_s} = e^{i\Delta} tan\Psi
\end{equation}

where $r_p$ and $r_s$ are the parallel and perpendicular reflection coefficients, respectively. In our experiment, these quantities are measured by two parameters, namely $I_s$ and $I_c$ as in Eq. (2) below. 

\begin{equation}
I_s = sin(2\Psi) sin(\Delta)~~and~~I_c = sin(2\Psi) cos(\Delta)
\end{equation} 

The refractive index (\emph{n}) and extinction coefficient (\emph{k}) of these films are determined by fitting $I_s$ and $I_c$ with the modified Forouhi-Bloomer dipersion relation. This relation fits smoothly for broader wavelength range \textit{i.e.} from normal to anomalous dispersion region \cite{Philipp}. This is also consistent with Kramers-Kronig relation with five independent parameters as described below.

\begin{equation}
\emph{n}(\omega)= \emph{n}_\infty + \sum_{j=1}^N \frac{B_j(\omega-\omega_j)+ C_j}{(\omega-\omega_j)^2+\Gamma_j^2}
\end{equation}

\begin{equation}
\emph{k}(\omega) = \left\{
  \begin{array}{lr}
    \sum_{j=1}^N \frac{f_j(\omega-\omega_g)^2}{(\omega-\omega_j)^2+\Gamma_j^2} & :\omega~\textgreater~\omega_g\\
    0 & ~~~~~:\omega\leq\omega_g
  \end{array}
\right.
\end{equation}
 
where, the term $f_j$ (in eV) is the oscillator strength, $\Gamma_j$ (in eV) is the broadening factor of absorption peak, $\omega_j$ (in eV) is the energy at which the extinction coefficient is maximum and $\omega_g$ (in eV) is the minimum energy from which absorption starts.

\begin{figure}
\centering
\includegraphics[width=0.70\textwidth]{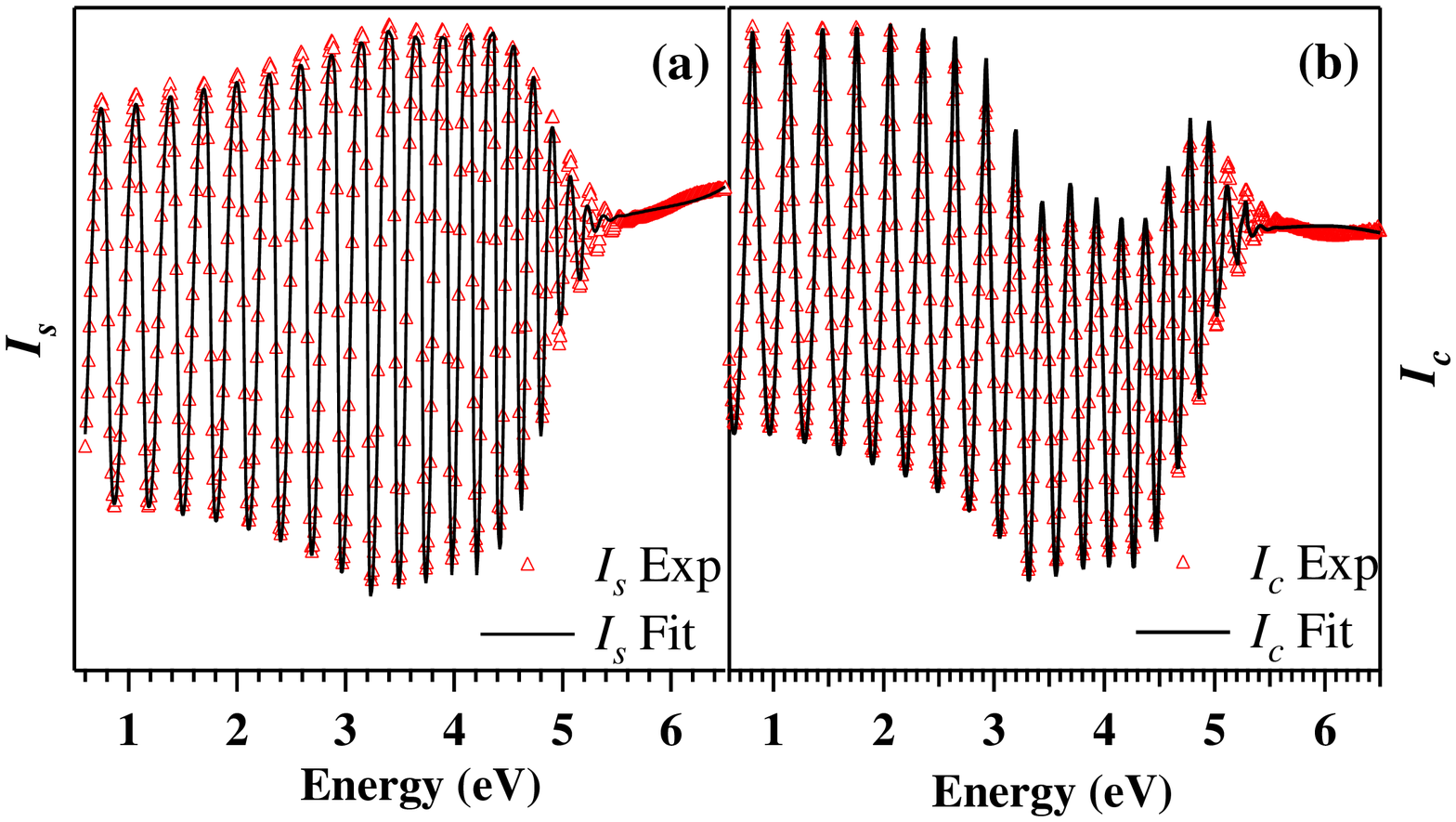}
\caption{(Color online) A plot of experimental parameters and corresponding fit as (a) $I_s$ and (b) $I_c$ for the AlN film grown at 400 $^\circ$C.}
\label{fig:2}
\end{figure}

In this study, the data analysis and fitting are performed using DeltaPsi2 software. A systematic approach of inclusion and omission of layers has been followed in which, a five layer model with  uniaxial anisotropic conditions (Air/roughness (AlN + void)/ AlN/interface (Al+AlN)/ Si) is employed for the analysis of optical properties of these AlN films. It is well known that wurtzite structure of AlN is a uniaxial anisotropic material, which exhibits directional dependent refractive index and warrants the application of an anisotropic model \cite{Jiang, Li, Shokhovets}. The fitting is performed by the classical non-linear minimization Levenberg-Marquardt algorithm as described by Modreanu \textit{et al} \cite{Modreanu}. Measured experimental parameters ($I_s$ and $I_c$) for AlN film grown at a typical $T_s$ of 400 $^\circ$C over a spectral range 0.6 to 6.5 eV and corresponding fit are shown in Fig.\ref{fig:2}. It is seen that the fitting of $I_s$ and $I_c$ across the whole spectral range is in good agreement.

\begin{figure}[h]
\centering
\includegraphics[width=0.50\textwidth]{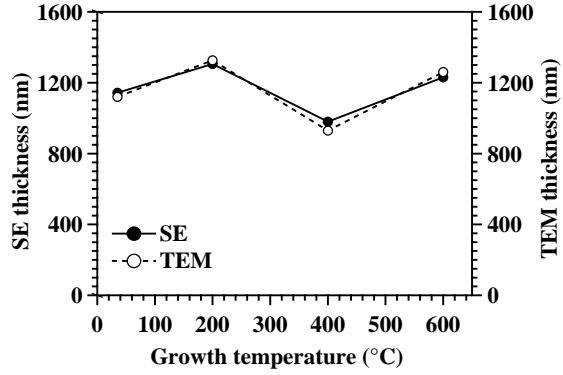}
\caption{Thickness of AlN films measured by SE and TEM.}
\label{fig:3}
\end{figure}

The nominal film thicknesses as computed from SE modeling compare well with those obtained from the x-TEM and is shown in Fig.\ref{fig:3}. The roughness as deduced from SE of these films is found to lie between 11 and 30~nm. These values are higher than the root mean square roughness values (3 to 9~nm) as obtained using Atomic Force Microscopy (AFM). The surface roughness obtained by AFM technique is acquired over an area of  1.5$\times 1.5~\mu m^{2}$ and thus basically represents the local roughness values of these films. In SE technique, the data are acquired over a larger elliptical area of 2$\times 0.7~mm^{2}$. Thus, the difference in the magnitude of roughness is due to the fact that the data for ellipsometry is from a larger area as compared to AFM \cite{Panda, Easwarakhanthan, Tripura}. Nevertheless, the roughness computed from SE and AFM both follows a similar trend with $T_s$.  The best fit with thin interface layer (Al+AlN) between AlN film and Si substrate is measured which decrease from 16 to 5 nm as function of $T_s$ with an Al volume fraction around 20 to 11\%.
 
\begin{figure}[h]
\centering
\includegraphics[width=0.70\textwidth]{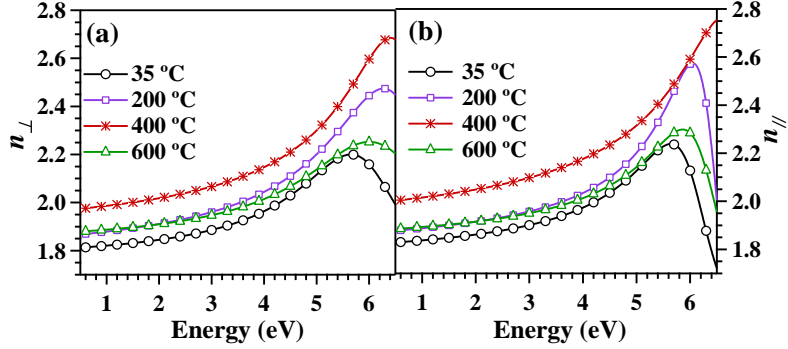}
\caption{(Color online) A plot of (a) $n_\bot$ and (b) $n_\|$ refractive index against to energy of AlN films for different $T_s$.}
\label{fig:4}
\end{figure}

\begin{figure}[h]
\centering
\includegraphics[width=0.70\textwidth]{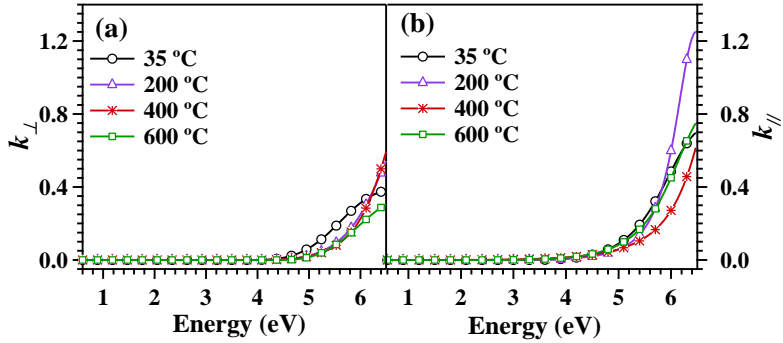}
\caption{(Color online) A plot of (a) $k_\bot$ and (b) $k_\|$ extinction coefficient against to energy of AlN films for different $T_s$.}
\label{fig:5}
\end{figure}

\begin{figure}[h]
\centering
\includegraphics[width=0.50\textwidth]{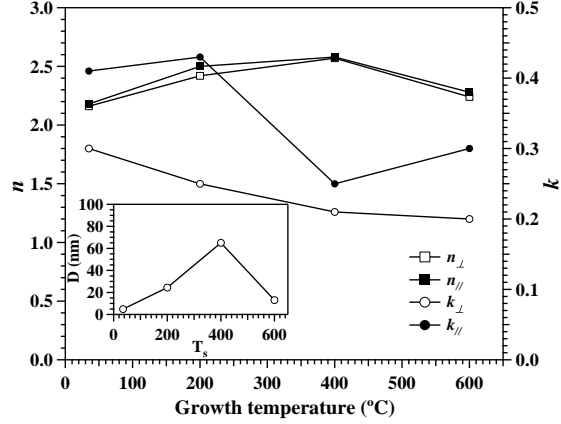}
\caption{The variation of \emph{n} and \emph{k} value at 210 nm and the crystallite size (inset) of AlN films with $T_s$.}
\label{fig:6}
\end{figure}

The real (\emph{n}) and imaginary (\emph{k}) parts of the refractive index of AlN films for different $T_s$ obtained from the five layer model are shown in Fig.\ref{fig:4} and Fig.\ref{fig:5}, respectively. The \emph{n} and \emph{k} exhibit strong uniaxial anisotropic dispersion and increase monotonically with increasing photon energy in normal dispersion region, while it decreases in anomalous region. Since, AlN is predominantly used in deep-UV region (210 nm), it is worthwhile to explore the behavior of both \emph{n} and \emph{k} of these films as a function of $T_s$ at 210 nm (5.9 eV) and are shown in Fig.\ref{fig:6}.

 With the increase in $T_s$, both $n_\bot$ and $n_\|$ increase up to 400 $^\circ$C and then thereafter it falls. While, both $k_\bot$ and $k_\|$ decreased with respect to $T_s$. All these films show transparent nature up to a photon energy 3.5 eV (354 nm) from NIR. As it is well known that the optical parameter \emph{n} and \emph{k} are basically dependent on the crystal structure, disorder like void, lattice defects and chemical composition of the film under investigation \cite{Zhao, G}. The crystal parameter such as crystallite size of these AlN films comparatively follows the behavior of refractive index. In the present study, the crystallite size of these films are calculated using Williamson-Hall method and is shown in the inset of Fig.\ref{fig:6}. It increases from 35 to 400 $^\circ$C and there after fall in crystallite size at 600 $^\circ$C due to dissociation of bonds and strong re-evaporation of ad-atoms at high $T_s$ \cite{Panda, Medjani, Kuan}. The parameter $T_s/T_m$ ($T_m$ = melting temperature of growing material) is important for the growth of film that defines the film orientation and structure \cite{Thornton}. If the film is deposited using sputtering at room temperature, there is a little surface diffusion due to high melting point of AlN and one would expect voids, nitrogen vacancy and defect concentration to be much higher than at equilibrium. But, with increase in $T_s$, the adatom mobility increases and causes the increase in crystalite size and columnar structure. The higher mobility of adatoms causes the formation of dense AlN films and reduces the residual stress, porosity and defects \cite{Panda}. Increase in growth or annealing temperature reduces the concentration of defect states like nitrogen, impurity and coordination defects, where refractive index of the film is improved and is proportional to packing density \cite{Zhao, G}. Hence, \emph{n} value is increasing linearly with $T_s$ upto 400 $^\circ$C, then a fall at 600~$^\circ$C due to decrease in crystallite size. Similarly, at 35 $^\circ$C the value \emph{k} is higher due to more vacancy in the film and Al concentration in interface layer, that can additionally contribute to the absorption by creating localized states. But, with increase in $T_s$, ad-atoms have high surface mobility that reduces defects and the Al concentration in the interface layer. That cause in reduce of \emph{k} value. Thus, the optical parameters \emph{n} and \emph{k} strongly depend on $T_s$ as well as crystallite size.

\begin{figure}[h]
\centering
\includegraphics[width=0.5\textwidth]{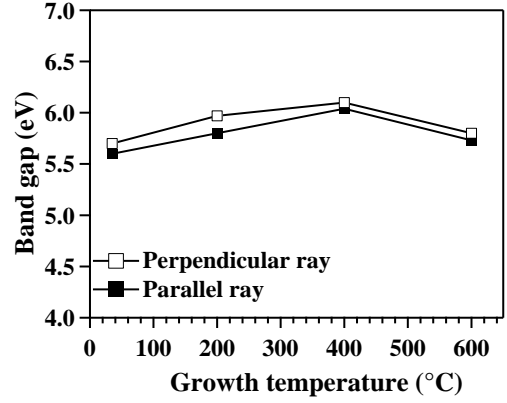}
\caption{The optical band gaps of AlN films as a function of $T_s$ for the perpendicular and parallel rays.}
\label{fig:7}
\end{figure}

To obtain the optical bandgap of these AlN films, the absorption coefficient ($\alpha$) defined as $\alpha$ = 4$\pi$\emph{k}/$\lambda$, where $\lambda$ is the wavelength of the incident light is calculated over extended energy range (0.6 to 6.5 eV) using the formalism followed by Looper~\emph{et al} \cite{Philipp}. The optical band gaps for the perpendicular and parallel rays are obtained using a linear extrapolation technique of tangential line to the energy axis of (E$\alpha$)$^2$, which is shown in Fig.\ref{fig:7}. These results agree well with the literature Kar~\emph{et al}, where they are reported the bandgap energy of AlN films after annealed as a function of temperature \cite{Kar}. The band gap increases with $T_s$ up to 400 $^\circ$C then it decreases at 600 $^\circ$C. So, optical band gap also strongly depends on the crystallite size. At low temperatures, the band gap is small compared to bulk AlN that is due to the generation of shallow states caused by the formation of lattice distortion by  voids, Al and N vacancy concentration \cite{Zhao, Kar}. Whereas, highly a-axis oriented AlN film grown at 400 $^\circ$C with large crystallite size ($\thicksim$ 66 nm) shows direct band gap as 6.05 and 6.1 eV, which is near to the bulk AlN.

\begin{figure}[h]
\centering
\includegraphics[width=0.7\textwidth]{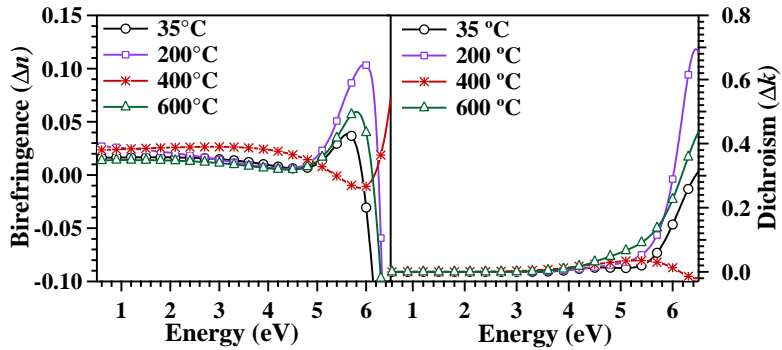}
\caption{(Color online) The dispersion of (a) birefringence ($\Delta n$) (b) and dichroism ($\Delta k$) with energy at different $T_s$.}
\label{fig:8}
\end{figure}

To undestand the behavior of anisotropy, the difference in $n$ (birefringence, $\Delta n$ = $n_\|$ - $n_\bot$) and $k$ (dichroism, $\Delta k$ = $k_\|$ - $k_\bot$) is shown in Fig.\ref{fig:8}. The $\Delta n$ exhibits a positive value except in higher energy region. Additionally, the value of $\Delta n$ is larger near band energy region for all these films, whereas AlN film grown at 400~$^\circ$C shows a maximum birefringence at lower energy range and a minimum nagative value at near band edge. Also, all these films exhibit a strong dichroism near the band edge, whereas 400~$^\circ$C AlN film shows a negative dichroism after the band edge. Generally, near the band gap, the fundamental absorption occurs due to the contribution of excitonic and band-to-band transitions. For wurzite structure AlN, the excitonic transition strongly depends upon polarization state of light due to the non-cubic crystal-field splittings \cite{Chen, Silveira}. Therefore, the anisotropy properties strongly affect near the band gap, which results a strong $\Delta n$. In this study, AlN grown at 400~$^\circ$C is a highly a-axis oriented along normal to the substrate compared to other films. Therefore, it shows a strong birefringence at lower energy range. Also, it is seen that A or M-plane LED shows isotropic emission pattern along the surface normal compared to the C-plane LED structure at 210 nm wavelength \cite{Taniyasu}. So, this film contains a mixed $\sigma$ and $\pi$ exciton feature, which decreases the value of both $\Delta n$ and $\Delta k$ near the band edge.

\begin{figure}[h]
\centering
\includegraphics[width=0.5\textwidth]{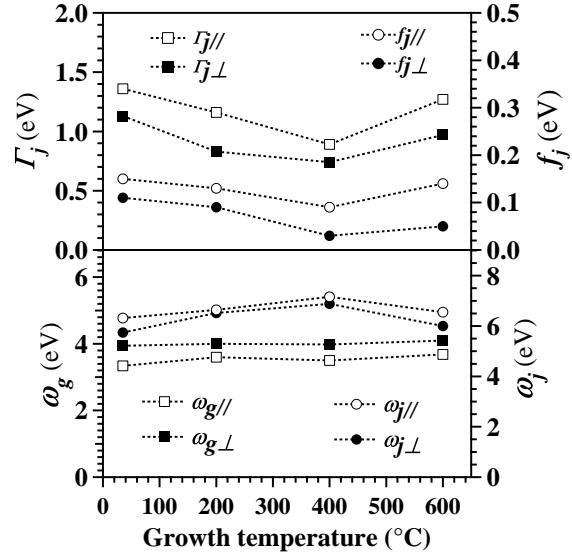}
\caption{Variation of dispersion parameters derived from the fitting against to $T_s$.}
\label{fig:9}
\end{figure}

The parameters of the modified Forouhi-Bloomer dipersion model were extracted from the fitting and shown in Fig. \ref{fig:9}. The $\Gamma_j$ and $\emph{f}_j$, both decrease with $T_s$ upto 400 $^\circ$C. The broadening parameter $\Gamma_j$ (= $\hbar$/$\tau_j$, where $\tau_j$ is the phonon relaxation time), is the inverse of relaxation time, depends on the phonon contribution and microstructural parameters, such as static impurities, defect density, strain, grain boundary, grain sizes, etc \cite{Sundari}. A decrease in $\Gamma_j$ is observed with the increase in $T_s$ and implies an increase in phonon relaxation time due to the increase in crystallite size as well as decrease in residual stress. $\omega_j$ defines the energy at which the extinction coefficient is maximum for a material and it increases with $T_s$ due to increase in band gap, whereas energy from which the absorption starts ($\omega_g$) is almost constant. So, at 400 $^\circ$C, AlN film shows high $n$ as well as low $k$ with higher value of $\omega_j$ among all due to higher purity, larger crystallite size and also highly a-axis oriented. 
  
Optical properties of AlN films with different $T_s$ by reactive DC magnetron sputtering are investigated by SE technique in the energy range of 0.6 to 6.5 eV. Thickness, roughness and optical constants of these films are obtained by the uniaxial anisotropic modeling with the modified Forouhi-Bloomer dipersion relation to fit the experimental SE data. The anisotropic optical parameters \emph{n} and \emph{k} are strongly depended on $T_s$. A highly a-axis oriented AlN film grown at 400 $^\circ$C, exhibited high \emph{n} ($n_\bot$ = 2.55, $n_\|$ = 2.57) and low \emph{k} ($K_\bot$ = 0.21, $K_\|$ = 0.25) at 210 nm (deep-UV region) with low value of $\Delta n$ and $\Delta k$. With increase in $T_s$, band gap also increased upto 400 $^\circ$C, which is close to the bulk AlN. So, the anisotropy optical properties of AlN are useful to UV-LED, polarizer, electro-luminescent diodes and AlN-based polarization-sensitive optoelectronic applications.

~

One of the authors (PP) acknowledges the research fellowship from the Department of Atomic Energy, Government of India.

\end{document}